\documentclass[twocolumn]{aastex631}
\usepackage[utf8]{inputenc}
\usepackage{amsmath}
\usepackage{amssymb}
\usepackage[centercolon]{mathtools}
\usepackage{color}
\usepackage{graphicx}
\usepackage{tikz}
\usepackage{mwe}
\usepackage[normalem]{ulem}
\usetikzlibrary{shapes.geometric, arrows}

\newcommand{\be}{\begin{equation}}
\newcommand{\ee}{\end{equation}}

\begin{document}

\title{Globally optimal and scalable $N$-way matching of astronomy catalogs}

\author{Tu Nguyen}
\affiliation{Dept. of Applied Mathematics \& Statistics, Johns Hopkins University, Baltimore MD 21218}

\author[0000-0002-1070-2626]{Amitabh Basu}
\affiliation{Dept. of Applied Mathematics \& Statistics, Johns Hopkins University, Baltimore MD 21218}
\affiliation{Dept. of Computer Science, Johns Hopkins University, Baltimore MD 21218}

\author[0000-0002-7034-4621]{Tam\'as Budav\'ari}
\affiliation{Dept. of Applied Mathematics \& Statistics, Johns Hopkins University, Baltimore MD 21218}
\affiliation{Dept. of Computer Science, Johns Hopkins University, Baltimore MD 21218}
\affiliation{Dept. of Physics \& Astronomy, Johns Hopkins University, Baltimore MD 21218}

\correspondingauthor{Tamas Budavari}
\email{budavari@jhu.edu}

\begin{abstract}
Building on previous Bayesian approaches, we introduce a novel formulation of probabilistic cross-identification, where detections are directly associated to (hypothesized) astronomical objects in a globally optimal way. 
We show that this new method scales better for processing multiple catalogs than enumerating all possible candidates, especially in the limit of crowded fields, which is the most challenging observational regime for new-generation astronomy experiments such as the Rubin Observatory Legacy Survey of Space and Time (LSST). 
Here we study simulated catalogs where the ground-truth is known and report on the statistical and computational performance of the method. 
The paper is accompanied by a public software tool to perform globally optimal catalog matching based on directional data.

\end{abstract}


\section{Motivation}
Several approaches have been proposed over the years to combine observations across telescopes and epochs.
In the era of time-domain astronomy where dozens or hundreds of observation epochs are available, these problems are more important than ever.
In particular, combining catalogs has been a central issue where detections in separate exposures are matched by identifying the ones that correspond to the same celestial object.
Several tools were developed to provide solution to the cross-matching problem, such as TOPCAT \citep{taylor2015} and CDS XMatch \citep{Pineau_2011, boch2012}. However, they do not consider the statistical aspect of the problem.

This cross-identification problem was successfully addressed using Bayesian hypothesis testing by \citet{budavari_szalay_2008}, whose methodology was implemented in the latest version of the SkyQuery service \citep{skyquery} which is now part of the SciServer Science Platform \citep{sciserver}. 
The Bayesian formalism and the combinatorial nature of the problem is discussed in a review by \citet{budavari_loredo_2015}.
%
The first solution came from \citet{budavari_basu_2016} who formulated the matching problem as a search for globally optimal associations using combinatorial optimization, where the marginal likelihood of the entire matched catalog is maximized, and used the Hungarian algorithm \citep{Munkres57hunalg} to solve it.
After that proof of concept was developed for two catalogs, \cite{shi2019probabilistic} extended the algorithm to handle multiple catalogs using Integer Linear Programming, or ILP for short.
For simplicity, we will refer this method as {\em{}CanILP}, as it enumerates all possible candidate associations and uses ILP to choose the best valid subset. As we will discuss later, the method suggested in~\cite{shi2019probabilistic} does not scale very well with large number of catalogs. This scaling problem is also observed in \citet{Pineau_2017} as the authors try to estimate the probability, for all combinations of sources, that a tuple of sources from different catalogs correspond to the same object. The exhaustive search results in an exponential growth in the number of possible tuples as the number of catalogs increases. They also note that this approach is not feasible in practice for more than 9 catalogs. 

In this paper, we improve on the previous studies by introducing a novel formulation, hereafter referred to as {\em DirILP}, where we use ILP to directly assign detections to hypothesized objects. 
Section~\ref{sec:theory} describes the new approach, and Section~\ref{sec:result_analysis} illustrates how the new method scales better with the number of input catalogs. Section~\ref{sec:software} discusses a public software tool to solve the catalog matching problem.  
Section~\ref{sec:future} concludes the study.

\section{Our Approach}\label{sec:theory}

To quantify the associations among independent detections, a relatively recent approach was developed that uses a hierarchical Bayesian formalism. 
Suppose there are $C$ catalogs, indexed by $c \in \{1, \ldots, C\}$, with each catalog capturing $N_c$ sources respectively. Let $D_{ic}$ denote the measurements for source $i$ in catalog $c$, hereafter denoted by tuple $(i,c)$.
Associated with any $(i,c)$ measurement is a likelihood function $\ell_{ic}(\omega) = p(D_{ic}|\omega),$ for the unknown true direction $\omega$, which captures the astrometric uncertainty. 
While other object properties could also be considered in general, such as their brightness or colors, here we focus on directional matching only.


%

Here we adopt the definition for matching used in previous papers \citep[e.g.,][]{budavari_szalay_2008, budavari_loredo_2015, budavari_basu_2016, shi2019probabilistic} where one tests whether two or more detections correspond to the same physical object. 
That said, it is possible to define the ``match'' hypothesis such that detections are not of the ``same'' object but instead just ``part of'' another, e.g., an optical galaxy in an X-ray cluster, or a star in a blend of two. 
In theory such scenarios can be accommodated (by changing the marginal likelihood calculations, see below), but the computation requirements might increase, and the interpretation of the matched catalog would be more difficult.%
\footnote{Further complications emerge when different blends are to be matched, in which case one considers whether the detections ``share'' components, e.g., a common star in two different blends.}
Our association approach described below is flexible and will yield matches based on the underlying model. 

Associations are created by grouping all sources in all catalogs such that each belongs to only one group. 
Mathematically, a \textit{partition} $P$ is created of the data set $D$, the union of all sources in all catalogs, where each subset corresponds to the same celestial object. 
The number of subsets in the partition will constitute the number of hypothesized objects $N_{\rm{}obj}$, which is an unknown but bounded quantity that is less or equal to the number of all sources. Typically it is much less as the equality would mean that every source is in fact a separate object altogether.
We can index every object by an integer \mbox{$o \in \{1, \ldots, N_{\rm{}obj}\}$}. 
Let $S_o$ be the set of sources $(i,c)$ associated with object $o$ and $C_o$ be the list of catalogs containing sources associated with object $o$. 
Following \citet{budavari_loredo_2015}, the likelihood of a partition $P$ of all sources, a collection of the $S_o$ subsets, will be a product of conditionally independent terms,
\begin{equation}
\mathcal{L}(P) \equiv p(D|P) = \prod_o \mathcal{M}_o,
\end{equation}
where the marginal likelihood $\mathcal{M}_o$ for the association corresponding to object $o$ is 
\begin{equation}
\mathcal{M}_o = \int d\omega\, \rho_{C_o}(\omega)\!\!\prod\limits_{(i,c) \in S_o}\ell_{ic}(\omega).
\end{equation}
Here $\rho_{C_o}(\omega)$ is the prior probability density function of the object direction producing sources within (the footprint of) every catalog in the set $C_o$.
This notation enables the treatment of catalogs with different sky coverage, whose angular selection function would enter the prior on the latent direction $\omega$.
Technically, the $\rho_{C_o}(\omega)$ function is not simply the angular selection function of the intersection area of the catalogs, because it also incorporates the astrometric uncertainty: there is a non-zero probability of observing a source within a given footprint even if its true direction is outside the field of view, but this effect is negligible if the field of view is large in comparison to its boundary blurred by the astrometric uncertainty, which is the case for typical observations and surveys, but would not apply, for example, if a catalog were an aggregation of disjoint sky patches comparable in size to the point-spread function.

Alternatively, one can define the marginal likelihood for a non-association hypothesis, which assumes that every source in $S_o$ is a separate object on its own
\begin{equation}
\mathcal{M}_o^{\rm{}N\!A} = \prod\limits_{(i,c) \in S_o}\int d\omega\, \rho_{c}(\omega)\,\ell_{ic}(\omega),
\end{equation}
where $\rho_{c}(\omega)$ is the prior probability density function of direction for sources in catalog $c$. 
This hypothesis serves as a natural comparison, with which it is useful to introduce the Bayes factor as the ratio of the marginal likelihoods of these two cases,
\begin{equation}
B_o = \frac{\mathcal{M}_o}{\mathcal{M}_o^{\rm{}N\!A}}.
\end{equation}
The $B_o$ takes values larger than 1 when the association of the sources in $S_o$ is more likely than the alternative, \mbox{$B_o<1$} favors separate objects. 
We note that \mbox{$\prod \mathcal{M}_o^{\rm{}N\!A}$} is simply a product over all sources in all catalogs independent of partition $P$ and, hence, is constant. Consequently, the maximization of the likelihood ${\cal{}L}(P)$ is equivalent to optimizing $\prod B_o$.

As customary, we work with a summary of the raw imaging data $D_{ic}$ for each detection $(i,c)$, the measured direction $x_{ic}$, which is essentially the intensity-weighted pixel direction.
In order to calculate the Bayes factors $B_o$ with these measurements, we specify a distribution for the member likelihood function $\ell_{ic}(\omega)$, i.e., the astrometric uncertainty. To describe directional uncertainty in the observations, the spherical analog of the Gaussian is often assumed, the \citet{Fisher1953} distribution,
\begin{equation}
\ell_{ic}(\omega) \coloneqq f(x_{ic};\omega, \kappa_{ic}) = \frac{\kappa_{ic}}{4\pi\sinh{\kappa_{ic}}}\exp\left({\kappa_{ic}\,\omega\cdot x_{ic}}\right),
\end{equation}
where the $x_{ic}$ observed direction and true $\omega$ direction are both 3D unit vectors. 
The latter is the mode of the distribution, and $\kappa_{ic}$ is a concentration parameter. 
When $\kappa_{ic} \gg 1,$ the Fisher distribution approximates a Gaussian distribution with standard deviation (in radians) for each coordinate $\sigma_{ic}$ with \mbox{$\kappa_{ic} = 1/\sigma_{ic}^2$} and the (all-sky) Bayes factor can be calculated analytically as shown in \cite{budavari_szalay_2008} as follows,
\begin{equation}\label{Bayes_factor}
B_o = 2^{\vert S_o \vert - 1} \frac{\prod_{ic} \kappa_{ic}}{\sum_{ic} \kappa_{ic}} \exp{\left(-\frac{\sum_{ic}\sum_{i'c'}\kappa_{ic}\kappa_{i'c'}\psi_{ic,i'c'}^2}{4\sum_{ic}\kappa_{ic}}\right)},
\end{equation}
where $(i,c)$ and $(i',c')$ are all sources in subset $S_o$ and $\psi_{ic,i'c'}$ is the (small) angle between the directions for sources $(i,c)$ and $(i',c')$.

The next section discusses how to find the globally optimal associations, i.e., the partition $P$ that maximizes the ${\cal{}L}(P)$ likelihood function by optimizing 
$\prod B_o$ using integer linear programming.

\subsection{CanILP: Optimal Selection of Candidates}\label{sec:CanILP_setup}

First we summarize the previous approach introduced by \citet{shi2019probabilistic} and highlight some of the outstanding challenges.
Maximizing $\prod B_o$ is equivalent to minimizing
\begin{equation}
-\sum_{o} \ln B_o.
\end{equation}
Given a data set $D$ of all $(i, c)$ pairs for all catalog $c$ and source $i$ in catalog $c$, we introduce a binary variable $x_T$ taking values in $\{0, 1\}$ for each nonempty subset $T \subseteq D$, with the interpretation that $x_T = 1$ indicates that the subset $T$ is included in the partition. To ensure the validity of the partition, we require
\be
\sum\limits_{T \ni (i,c)} x_T = 1
\ee
for every element $(i,c) \in D.$ This forces every source $(i,c)$ to be included in exactly one subset of the partition. However, note that for an orphan $o,$ $B_o = 1.$ Hence, these coefficients do not contribute to the objective function and we could simply remove those subsets $T$ that have $\lvert T \rvert = 1.$ From this, we can modify the above constraint to 
\be
\sum\limits_{T \ni (i,c)} x_T \leq 1
\ee
for every element $(i,c) \in D.$ In the final solution, if a source $(i,c)$ does not appear in any subset $T,$ we treat it as an orphan. For example, in Figure \ref{fig:CanILP_eg}, \texttt{Source (2,1)} is not included in any subset $T$, so, in the solution, we will include it as an orphan. 

By defining $w_T = -\ln B_T$ for every subset $T$, the final integer linear programming function can be written as follows,
\begin{gather} 
\min \sum_T w_T x_T \nonumber \\
\text{subject to } x_T \in \mathbb{Z} \text{ and } 0 \leq x_T \leq 1 \text{ for all } T, \nonumber \\
\text{and } \sum_{T \ni (i,c)} x_T \leq 1 \text{ for all } (i,c) \in D.
\end{gather}
%
%
Note that the formulation above can be used to solve the matching problem given any number of catalogs $C$ but requires an enumeration of all possible candidate associations, which can be numerous. This requires calculations of combinatorially many $w_T$ and the introduction of the corresponding $x_T$ binary variables, which quickly becomes prohibitively expensive for many catalogs.

\cite{shi2019probabilistic} demonstrated their approach on three catalogs with identical astrometric uncertainty.
%
Here we extend their work and describe a novel approach, which can be efficiently used with many catalogs. 

\tikzstyle{catalog} = [rectangle, rounded corners, minimum width=3cm, minimum height=1cm,text centered, draw=black, fill=red!30]
\tikzstyle{source} = [circle, text centered, draw=black, fill=orange!30]
\tikzstyle{object} = [circle, text centered, draw=black, fill=blue!30]
\tikzstyle{assoc} = [diamond, minimum width=3cm, minimum height=1cm, text centered, draw=black, fill=white!30]
\tikzstyle{activate} = [diamond, minimum width=3cm, minimum height=1cm, text centered, draw=black, fill=green!30]
\tikzstyle{arrow} = [thick,->,>=stealth]

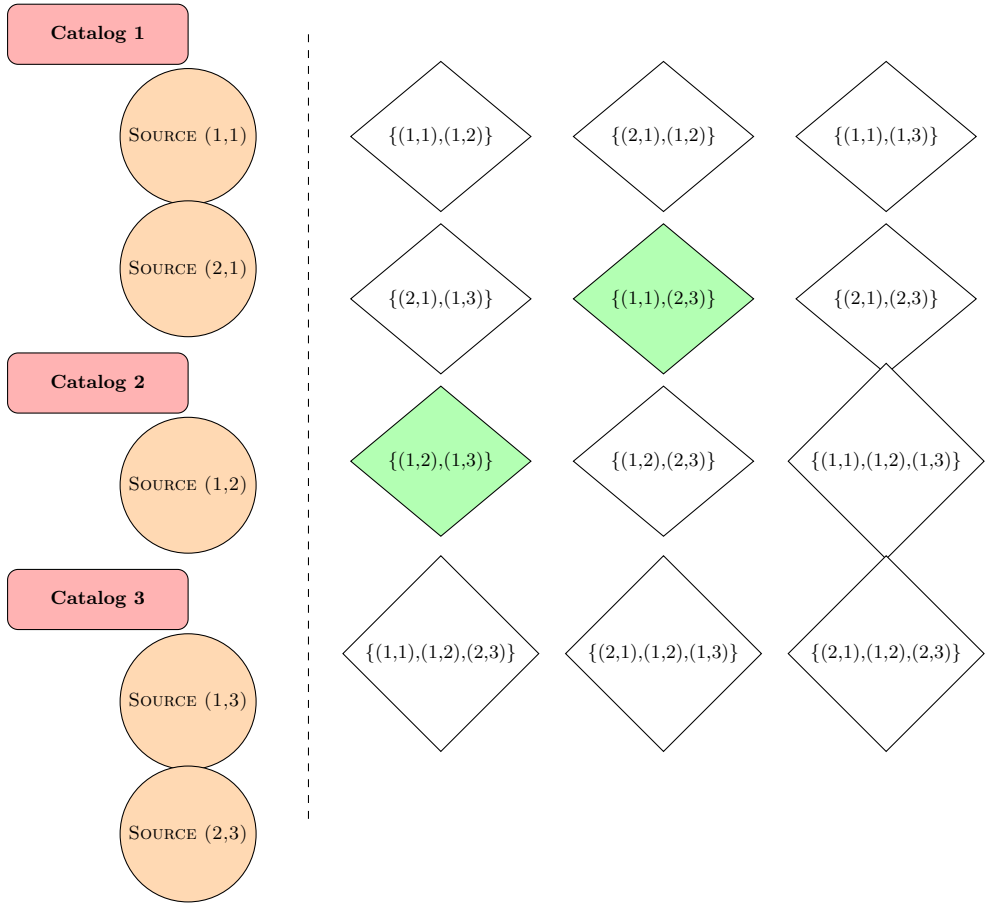
\begin{figure*}
\centering
\begin{tikzpicture}[node distance=1.7cm, every node/.style={scale=0.8}]
\node (catalog_1) [catalog] {\textsc{\textbf{Catalog 1}}};
\node (source_11) [source, below of=catalog_1, xshift = 1.5cm] {\textsc{Source (1,1)}};
\node (source_21) [source, below of=source_11, yshift = -0.5cm] {\textsc{Source (2,1)}};
\node (catalog_2) [catalog, below of=source_21, xshift = -1.5cm, yshift=-0.2cm]  {\textsc{\textbf{Catalog 2}}};
\node (source_12) [source, below of=catalog_2, xshift = 1.5cm] {\textsc{Source (1,2)}};
\node (catalog_3) [catalog, below of=source_12, xshift = -1.5cm, yshift=-0.2cm]  {\textsc{\textbf{Catalog 3}}};
\node (source_13) [source, below of=catalog_3, xshift = 1.5cm] {\textsc{Source (1,3)}};
\node (source_23) [source, below of=source_13, yshift = -0.5cm] {\textsc{Source (2,3)}};
\node (assoc_1) [assoc, right of=source_11, xshift = 2.5cm] {\{(1,1),(1,2)\}};
\node (assoc_2) [assoc, right of=assoc_1, xshift = 2cm] {\{(2,1),(1,2)\}};
\node (assoc_3) [assoc, right of=assoc_2, xshift = 2cm] {\{(1,1),(1,3)\}};
\node (assoc_4) [assoc, below of=assoc_1, yshift = -1cm] {\{(2,1),(1,3)\}};
\node (assoc_5) [activate, below of=assoc_2, yshift = -1cm] {\{(1,1),(2,3)\}};
\node (assoc_6) [assoc, below of=assoc_3, yshift = -1cm] {\{(2,1),(2,3)\}};
\node (assoc_7) [activate, below of=assoc_4, yshift = -1cm] {\{(1,2),(1,3)\}};
\node (assoc_8) [assoc, below of=assoc_5, yshift = -1cm] {\{(1,2),(2,3)\}};
\node (assoc_9) [assoc, below of=assoc_6, yshift = -1cm] {\{(1,1),(1,2),(1,3)\}};
\node (assoc_10) [assoc, below of=assoc_7, yshift = -1.5cm] {\{(1,1),(1,2),(2,3)\}};
\node (assoc_11) [assoc, below of=assoc_8, yshift = -1.5cm] {\{(2,1),(1,2),(1,3)\}};
\node (assoc_12) [assoc, below of=assoc_9, yshift = -1.5cm] {\{(2,1),(1,2),(2,3)\}};
\draw [dashed] (2.8,0) -- (2.8,-10.5);
\end{tikzpicture}
\caption{An illustration of CanILP. As can be seen on the left side, we assume there are $2$ detections from Catalog 1 (\texttt{Sources (1,1)} and \texttt{(2,1)}), $1$ detection from Catalog 2 (\texttt{Source (1,2)}) and $2$ detections from Catalog 3 (\texttt{Sources (1,3)} and \texttt{(2,3)}). In CanILP, we list all candidates for possible associations across independent detections, which are shown on the right side. These are the $x_T$ in the formulation. We then find the combinations of subsets that maximize the overall likelihood. Here, the solution given by CanILP indicates that the subsets $\{(1,1),(2,3)\}$ and $\{(1,2),(1,3)\}$ are included in the partition. These subsets, which are represented by a green color, correspond to the variables $x_{\{(1,1),(2,3)\}} = x_{\{(1,2),(1,3)\}} = 1$ in the model. On the other hand, all other variables $x_T = 0.$ Notice that because \texttt{Source (2,1)} does not appear in any of these subsets, so we treat it as an orphan. As a result, the association outputted by CanILP is $\{\{(1,1),(2,3)\}, \{(1,2),(1,3)\}, \{(2,1)\}\}$.}
\label{fig:CanILP_eg}
\end{figure*}

\subsection{DirILP: Optimal Direct Associations}\label{sec:DirILP_setup}

The key idea is to introduce variables that directly assign the detections to hypothesized objects instead of simply switching on and off some previously enumerated candidate associations in the final matched catalog. 
The objective $\prod B_o$ is the same but expressing it with the new variables is significantly more complicated than before.
In the process one needs to introduce several additional (sets of) auxiliary variables to linearize the problem.
In case of homoscedasticity when all astrometric uncertainty are the same for all detections, the linearization is relatively straightforward, but further modeling tricks are required in the general setting.  
In the following sections these two cases are introduced along with the variables needed to model and solve the global association problem. 
Further details are provided in the appendix about the derivation of the general heteroscedastic formalism.

\subsubsection{Homoscedasticity}
For simplicity, we first discuss the special case where the astrometric uncertainty of each detection is the same, i.e., \mbox{$\sigma_{ic}\!=\!\sigma$} for each source $(i,c)$. 
%
Given a data set $D$, let $N$ be the total number of detections in all catalogs considered. The number of astronomical objects these represent will be at most $N$, corresponding to the hypothesis that every detection comes from a different object.
%
Our goal is to find a mapping that matches each source to one (and only one) object. 
This association between a source and an object means that the source is an observation of that object in the sky. 
Naturally, multiple sources are expected to be assigned to the same object, which represents the hypothesis that all of these sources are observations of that same object. 
To capture the matching between a source $(i,c)$ and an object $o$, we introduce binary variables  
$\{x_{ic}^o\}$, 
where a given $x_{ic}^o=1$ if the $(i,c)$ detection is associated with object $o$, and 0 otherwise.

\begin{figure}
\centering
\begin{tikzpicture}[node distance=2cm, every node/.style={scale=0.7}]
\node (catalog_1) [catalog] {\textsc{\textbf{Catalog 1}}};
\node (source_11) [source, below of=catalog_1, xshift = 2cm] {\textsc{Source (1,1)}};
\node (source_21) [source, below of=source_11, yshift = -0.5cm] {\textsc{Source (2,1)}};
\node (catalog_2) [catalog, below of=source_21, xshift = -2cm] {\textsc{\textbf{Catalog 2}}};
\node (source_12) [source, below of=catalog_2, xshift = 2cm] {\textsc{Source (1,2)}};
\node (catalog_3) [catalog, below of=source_12, xshift = -2cm] {\textsc{\textbf{Catalog 3}}};
\node (source_13) [source, below of=catalog_3, xshift = 2cm] {\textsc{Source (1,3)}};
\node (source_23) [source, below of=source_13, yshift = -0.5cm] {\textsc{Source (2,3)}};
\node (object_1) [object, right of=source_21, xshift = 3cm, yshift = 2cm] {\textsc{Object 1}};
\node (object_2) [object, below of=object_1, yshift = -0.5cm] {\textsc{Object 2}};
\node (object_3) [object, below of=object_2, yshift = -0.5cm] {\textsc{Object 3}};
\node (object_4) [object, below of=object_3, yshift = -0.5cm] {\textsc{Object 4}};
\node (object_5) [object, below of=object_4, yshift = -0.5cm] {\textsc{Object 5}};
\draw [arrow] (source_11) -- (object_4);
\draw [arrow] (source_21) -- (object_1);
\draw [arrow] (source_12) -- (object_3);
\draw [arrow] (source_13) -- (object_3);
\draw [arrow] (source_23) -- (object_4);
\end{tikzpicture}
\caption{An illustration of DirILP. As in Figure \ref{fig:CanILP_eg}, assume there are $2$ detections from Catalog 1 (\texttt{Sources (1,1)} and \texttt{(2,1)}), $1$ detection from Catalog 2 (\texttt{Source (1,2)}) and $2$ detections from Catalog 3 (\texttt{Sources (1,3)} and \texttt{(2,3)}). In this case, the output of DirILP indicates that \texttt{Sources (1,1)} and \texttt{(2,3)} belong to the same object, that \texttt{Sources (1,2)} and \texttt{(1,3)} belong to the same object, and that \texttt{Source (2,1)} is an orphan. Notice that it is okay for an object to not have any source associated with it. The solution given by DirILP is $\{\{(1,1),(2,3)\}, \{(1,2),(1,3)\}, \{(2,1)\}\}$, which is the same as the one given by CanILP in Figure \ref{fig:CanILP_eg}.}
\label{fig:DirILP_eg}
\end{figure}
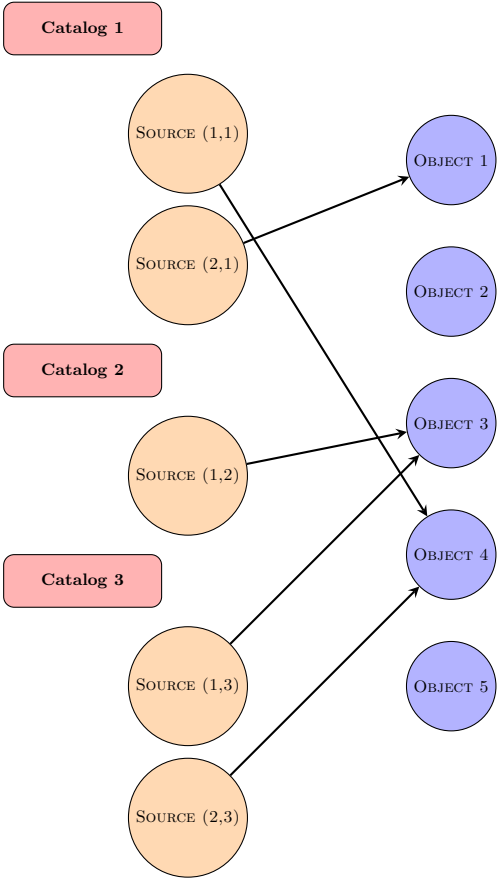

Figure~\ref{fig:DirILP_eg} illustrates how this approach works. For example, the arrow from \texttt{Source (2,1)} to \texttt{Object 1} representing an association means that $x_{21}^1 = 1$. Similarly, $x_{11}^3 = 0$ means no association, hence there is no arrow between the corresponding entries.
A partition $P$ can now be represented as a set 
\mbox{$\{S_o: o = 1,\dots, N\}$}, 
where $S_o$ is the subset of sources assigned to $o$, i.e., 
\begin{equation}
    S_o \coloneqq  \left\{(i,c): x^o_{ic} = 1\right\}.
\end{equation} 
If, for a given index $o$, \mbox{$x^o_{ic} = 0$} for all $(i,c)$ sources, then \mbox{$S_o = \emptyset$} is empty, which means object $o$ is not needed for that particular partition. 

Recall that the goal is to maximize the product of Bayes factors $\prod B_o$ (or to minimize $-\sum \ln B_o$) corresponding to these associations. Given an association $S_o$, assuming $\kappa_{ic} = \kappa$ for all source $(i,c)$, eq.~\eqref{Bayes_factor} gives us
\begin{align}
B_o &= 2^{\vert S_o \vert - 1} \frac{\prod_{ic} \kappa}{\sum_{ic} \kappa} \exp{\left(-\frac{\sum_{ic}\sum_{i'c'}\kappa^2\psi_{ic,i'c'}^2}{4\sum_{ic}\kappa}\right)} \\
& = 2^{\vert S_o \vert - 1} \frac{\kappa^{\vert S_o \vert}}{\vert S_o \vert \kappa} \exp{\left(-\frac{\kappa\sum_{ic}\sum_{i'c'}\psi_{ic,i'c'}^2}{4\vert S_o \vert}\right)}
\end{align}
Hence, 
\iffalse
\begin{align}
-\ln B_o &= (1 - \vert S_o\vert)\ln2 - \vert S_o \vert \ln\kappa + \ln \vert S_o\vert + \ln\kappa + \frac{\sum
\limits_{ic,i'c'}\kappa\psi_{ic,i'c'}^2}{4\vert S_o\vert} \\
& = \ln(2\kappa)(1 - \vert S_o\vert) + \ln \vert S_o\vert + \frac{\sum
\limits_{ic,i'c'}\kappa\psi_{ic,i'c'}^2}{4\vert S_o\vert}
\end{align}
\else
\begin{equation}
-\ln B_o = \ln(2\kappa) \left(1 - \vert S_o\vert\right) + \ln \vert S_o\vert + \frac{\kappa\sum
\psi_{ic,i'c'}^2}{4\vert S_o\vert}
\end{equation}
\fi
We want to find the partition $P$ that minimizes $- \sum \ln B_o.$ Notice that as of now, there are still several non-linear terms in $-\ln B_o$ so it is not yet a linear objective. To make use of ILP method, we will first need to rewrite this as a linear function. To do that, we introduce the following variables, defined for each index $k \in \{0, \ldots, C\}$, with $C$ representing the total number of catalogs:
\begin{equation}
z_k^o =
\begin{cases*}
      1 & if $\sum\limits_{ic} x_{ic}^o = k$ \\
      0       & otherwise
    \end{cases*}
\end{equation}
This variable captures the number of sources getting matched to object $o$, or the cardinality of the subset $S_o.$ When $z_{k'}^{o'} = 1,$ there are $k'$ hypothesized observations of object $o'$ in the data. In addition, notice that at most $1$ of the $z_k^o$, $k =0, \ldots, C$ can be $1.$ We also introduce

\begin{equation}
y_{ic, i'c'}^o =
\begin{cases*}
    1 & if $x_{ic}^o = x_{i'c'}^o = 1$  \\
    0 & otherwise
\end{cases*}
\end{equation}
This is an indicator variable that checks whether the sources $(i,c)$ and $(i',c')$ belong to the same object $o$. In particular, $y_{ic, i'c'}^{o'} = 1$ indicates the hypothesis that sources $(i,c)$ and $(i',c')$ are observations of object $o'.$ We also have

\begin{equation}
t^o =
\begin{cases*}
  \frac{\sum\kappa\psi_{ic,i'c'}^2 y_{ic, i'c'}^o}{4k} & if $z_k^o = 1$ for some $k \in [C]$ \\
  0 & if $z_0^o = 1$
\end{cases*}
\end{equation}
where $[C]$ represents the set of numbers $\{1, 2, \cdots, C \}$. This variable captures the last term in $-\ln B_o$ for a subset $S_o.$ In particular, when $z^o_k = 1$ for some $k \in [C]$, i.e. $\lvert S_o \rvert = k$ by definition of $z^o_k$, we have 
\begin{equation}
t^o = \frac{\sum
\kappa\,\psi_{ic,i'c'}^2}{4\vert S_o\vert}
\end{equation} 
as desired, where the summation goes over all $(i,c)$ and $(i',c')$ in $S_o.$ On the other hand, when $z^o_0 = 1,$ no detection is assigned to object $o$, so this term should contribute nothing to the objective function. 
Next, we introduce
\begin{equation}\label{eq:po}
p^o =
\begin{cases*}
  \left(1 - k\right) \ln(2\kappa) & if $z_k^o = 1$ for some $k \in [C]$ \;\; \\
  0       & if $z_0^o = 1$
\end{cases*}.
\end{equation}
This variable captures the first term in $-\ln B_o$ for a subset $S_o.$ It plays a similar role as $t^o$, i.e., when $z^o_0 = 1,$ no detection is assigned to object $o$, so this term should contribute nothing to the objective function. On the other hand, if some sources are matched to object $o$, $p^o = \ln(2\kappa)(1 - \vert S_o\vert)$ as desired.

Finally, we will linearize the term $\ln \lvert S_o \rvert$ by breaking the natural log function into finitely many affine linear pieces. We first introduce constants $a_1, a_2, \cdots, a_C,$ where $a_1 = 0$ and $a_p = \ln(p) - \ln(p-1)$, for \mbox{$p = 2, \cdots, C$}. Then for each object $o$, we define binary variables \mbox{$w_1^o \geq w_2^o \geq \cdots \geq w^o_C$} and impose the constraint that 
\begin{equation}
\sum_{p=1}^C w_p^o = \sum_{ic} x_{ic}^o = \lvert S_o \rvert.
\end{equation}
Using the new notation, we can now express $\ln \lvert S_o \rvert$ as a linear function of $w_p^{o}$: $\ln \vert S_o\vert = \sum_{p=1}^C a_p w_p^{o}$. To explain why this is the case, it is best to work with an example. Suppose 3 sources are matched with object $o,$ so $\lvert S_o \rvert = 3$ and $\ln \lvert S_o \rvert = \ln{3}.$ Because $\sum_{p=1}^C w_p^o = \vert S_o\vert = 3$ and $w_p^o$ are $0/1$ variables with $w_1^o \geq w_2^o \geq \cdots \geq w^o_C$, we have $w_1^o = w_2^o = w_3^o = 1$ and $w_4^o = w_5^o = \cdots = w_C^o = 0.$
Then, $\sum_{p=1}^C a_p w_p^{o} = a_1 + a_2 + a_3 = (0) + (\ln{2} - \ln{1}) + (\ln{3} - \ln{2}) = \ln{3},$ which is exactly $\ln \lvert S_o \rvert.$
Our objective function now becomes \begin{equation}\min \sum_o \bigg(\;\;p^o + \sum_p a_p w_p^o + t^o \bigg),\end{equation} which is linear in the variables $p^o,w^o_p$ and $t^o$.

As can be seen in the definitions of these variables, there are certain relationships that still need to be modeled using linear constraints. The full ILP formulation is given in Appendix~\ref{appx:DirILP_special} with detailed explanations for how the constraints model the relationships between the variables $x^o_{ic}, y^o_{ic,i'c'}, z^o_k, p^o, w^o_p,$ and $t^o$.

\subsubsection{Heteroscedasticity} 
\label{sec:DirILP_general}
We can also remove the assumption that every source has the same measure of uncertainty $\kappa_{ic}$.
From eq.~\eqref{Bayes_factor}, we have,
\iffalse
\begin{equation}\label{eq:ln_bayes_general} -\ln B_o = (1 - \vert S_o \vert) \ln 2 - \sum_{ic} \ln \kappa_{ic} + \ln \sum_{ic} \kappa_{ic} + \frac{\sum_{ic}\sum_{i'c'}\kappa_{ic}\kappa_{i'c'}\psi_{ic,i'c'}^2}{4\sum_{ic}\kappa_{ic}}, 
\end{equation}
\else
\begin{eqnarray}\label{eq:ln_bayes_general} -\ln B_o &=& (1 - \vert S_o \vert) \ln 2 - \sum_{ic} \ln \kappa_{ic} + \ln \sum_{ic} \kappa_{ic} + \nonumber \\ &+&\frac{\sum_{ic}\sum_{i'c'}\kappa_{ic}\kappa_{i'c'}\psi_{ic,i'c'}^2}{4\sum_{ic}\kappa_{ic}}, 
\end{eqnarray}
\fi
where all the summations run over all $(i,c)$ and $(i',c')$ in $S_o.$
We use $x_{ic}^o, z_k^o,$ and $y_{ic, i'c'}^o$ as defined in the special case of Section~\ref{sec:DirILP_setup}. We also introduce new variables to convert eq.~\eqref{eq:ln_bayes_general} into a linear function.

We first linearize the term $\ln\sum \kappa_{ic}$ using the same trick as when we linearized $\ln \lvert S_o \rvert$ in Section \ref{sec:DirILP_setup}. 
We introduce constants $b_{\min} \equiv b_1, b_2, b_3, \cdots$, where
\begin{equation}
b_{\min} = \ln\left(\min_{ic \in D} \kappa_{ic}\right)
\end{equation}
and 
\begin{equation}
b_{\max} = \ln\left(C \, \max_{ic \in D} \kappa_{ic}\right)\,.
\end{equation}
Now, if we set an error threshold $\epsilon,$ then the 
\begin{equation}
R \equiv \left\lceil \frac{b_{\max} - b_{\min}}{\epsilon} \right\rceil
\end{equation}
constants $b_p$ are defined as 
\begin{equation}
b_p = b_{\min} + (p-1) \times \epsilon \quad \textrm{for} \quad p = 1, \dots, R.
\end{equation}
Then for each object $o$, we define binary variables \mbox{$\chi_1^o \geq \chi_2^o \geq \cdots \geq \chi^o_P$} and impose the constraint 
\begin{equation}
\chi_1^o \exp(b_1) + \sum_{p=2}^R \chi_p^o \left[\exp(b_p) - \exp(b_{p-1})\right] \geq  \sum_{ic} \kappa_{ic} x_{ic}^o  \,.
\end{equation}
Using the new variables, we have 
\begin{equation}
\ln \sum_{ic} \kappa_{ic} \approx \chi_1^o b_1 + \sum_{p=2}^R \chi_p^o (b_p - b_{p-1}) = \chi_1^o\,b_{\min} + \epsilon \sum_{p=2}^R \chi_p^o
\end{equation}
since $b_p - b_{p-1} = \epsilon$ for all $p \geq 2$.

To illustrate how the $\chi_p^o$ variables work, let us assume that after looking at the data, we determine that \mbox{$b_{\min} = 29$} and \mbox{$b_{\max} = 33$}. If we let \mbox{$\epsilon = 0.5$}, then the value of constants $b_p$ are $\{29, 29.5, \cdots, 32.5, 33\}.$ Now suppose there are $3$ sources that are matched to an object $o$ with associated $\kappa_{ic}$ values of $5\times10^{12}, 8\times10^{12},$ and $10^{13}.$ Then the true value of $\ln \sum_{ic \in S_o} \kappa_{ic}$ is $\ln(2.3\times10^{13}),$ which evaluates to $30.77.$ With the defined variables, the solution given by ILP is $\chi_1^o = \chi_2^o = \cdots = \chi_5^o = 1$ and $\chi_6^o = \cdots = \chi_9^o = 0$ because $\chi_1^o \exp(b_1) + \sum_{p=2}^P \chi_p^o (\exp(b_p) - \exp(b_{p-1})) = \exp(29) + \exp(29.5) - \exp(29) + \exp(30) - \exp(29.5) + \cdots + \exp(31) - \exp(30.5) = \exp(31) > 2.3\times10^{13},$ which satisfies the constraint \mbox{$\chi_1^o \exp(b_1) + \sum_{p=2}^P \chi_p^o (\exp(b_p) - \exp(b_{p-1})) \geq  \sum_{ic} \kappa_{ic} x_{ic}^o.$} Notice that setting the variables $\chi_6^o, \cdots, \chi_9^o = 1$ will also satisfy the constraint. However, since we will model our problem with a minimization objective, the optimal solution will force $\chi_1^o b_1 + \sum_{p=2}^R \chi_p^o (b_p - b_{p-1})$ to be as small as possible. Finally, notice that in this case the value of $\chi_1^o b_1 + \sum_{p=2}^R \chi_p^o (b_p - b_{p-1})$, which is used to approximate $\ln \sum_{ic \in S_o} \kappa_{ic}$, is $31,$ which is close to the true value of $30.77.$ 

Next, we will linearize the last term in  eq.~\eqref{eq:ln_bayes_general} by first introducing the constant 
\be
c_{\min} = \min_{ic \in D} \kappa_{ic}
\ee
and 
\be
c_{\max} = C\, \max_{ic \in D} \kappa_{ic} \,.
\ee 
Then by rounding these two values to the nearest 100, we can introduce grid points $0 \equiv c_0, c_1, c_2, \cdots, c_Q,$ where $c_1$ is the nearest 100 of $c_{\min}$, $c_Q$ is the nearest 100 of $c_{\max}$, and for all $i > 2$, $c_i = c_1 + 100(i-1).$ We then introduce 

\begin{equation}
u_k^o = 
\begin{cases*}
  1 & if $\sum\limits_{ic} \left[ \kappa_{ic} \right]_{_{100}}\,  x_{ic}^o = c_k $ \\
  0       & otherwise
\end{cases*}
\end{equation}
where $k$ ranges in $\{0, 1, \ldots, Q\}$ and the operator $[\cdot]_{_{100}}$" is defined as rounding to the nearest $100$. This variable attempts to approximate $\sum_{ic \in S_o} \kappa_{ic}$, which appears in the denominator of the last term of eq.~\eqref{eq:ln_bayes_general}. 
The variables $p^o$ and $t^o$ are also very similar to the definitions in Section~\ref{sec:DirILP_setup}; however, we need to slightly modify them as follows:
%
%
\begin{equation}\label{eq:to-def}t^o = \frac{\sum_{ic}\sum_{i'c'}\kappa_{ic}\kappa_{i'c'}\psi_{ic,i'c'}^2 y_{ic, i'c'}^o}{4c_k},\end{equation} if $u_k^o = 1$ for some $k \in \{1, 2, \cdots, Q \},$ and $t^o = 0$ otherwise.

The reasoning for defining $t^o$ this way is that if \mbox{$u_0^o = 1$},
\be
\sum\limits_{(i,c)} [\kappa_{ic}]_{_{100}}x_{ic}^o = c_0 = 0.
\ee
This happens only when $x_{ic}^o = 0$ for all $(i,c)$, i.e. no sources are matched to object $o.$ Hence, $t^o$ should not contribute to the objective function, hence the value of $0$. On the other hand, if $u_k^o = 1$ for some $k > 0,$ by definition of $u_k,$ $c_k$ is the best approximation to $\sum_{ic \in S_o} \kappa_{ic}$. Thus,~\eqref{eq:to-def} holds.

In addition, we modify $p^o$ defined in eq.~\eqref{eq:po} as follows,
\begin{equation}
p^o =   
\begin{cases*}
  \left(1 - k\right) \ln(2) & if $z_k^o = 1$ for some $k \in [C]$ \;\; \\
  0       & if $z_0^o = 1$
\end{cases*}.
\end{equation}
This variable serves a similar function as in the homoscedastic case, which is to capture the first term in eq.~\eqref{eq:ln_bayes_general}. 

%

The objective function can now be written as \begin{equation}\sum_o \left(\;\;p^o - \sum_{ic}x^o_{ic}\ln\kappa_{ic} + \chi_1^o b_{\min} + \epsilon\sum_{p=2}^P \chi_p^o + t^o \right),\end{equation}
%
%
%
which is linear in all the variables involved.

There are certain relationships that still need to be modeled using linear constraints because ILP formulations only take linear constraints. The full ILP formulation is given in Appendix \ref{appx:DirILP_general}, with detailed explanations for how the constraints model the relationships between the variables $x^o_{ic}, y^o_{ic,i'c'}, z^o_k, \chi^o_p, u^o_k, p^o,$ and $t^o$.

\section{Mock Objects and Simulations}\label{sec:result_analysis}

We consider the idealized case where all the catalogs capture the same astronomical properties of objects in the sky, i.e., they detect the same set of objects. As we generate 100 objects and assume there are $C$ distinct catalogs, we expect to see $100 \times C$ sources and $100$ $C-$way association sets. We will now show the catalog matching results using both of our approaches. The ILP programs in both approaches are solved using Gurobi, an optimization solver \cite{gurobi}. 

\subsection{Homoscedasticity: identical \mbox{$\kappa_{ic} = 1/\sigma^2$}} 
\label{subsec:Homoscedasticity}

Observe that for the CanILP formulation in Section \ref{sec:CanILP_setup}, we need to list all the possible valid subsets $T \subseteq D.$ We could do this by sequentially adding catalogs one by one and considering sources from the new catalog. However, this evaluates to $101^C - 1$ subsets, which is exponential in terms of the number of catalogs $C.$ Hence, we first try to reduce the number of possible subsets by observing that sources that are far away cannot be from the same object. So we can impose some distance constraints on the sources that are put into the same candidate association set. In doing so, we should be careful not to discard potentially true associations later on because say two sources from the first $2$ catalogs that are far away might not be a $2-$way matching; but, if on the third catalog, there is a source lying in the middle of the path between these $2$ sources, the $3$ sources together might be a $3-$way matching. 

\begin{figure}[ht]
    \centering
    \includegraphics[width=0.48\textwidth]{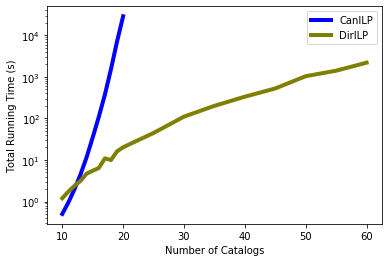}
    \caption{Total running time comparison between the two formulations for the special case (Log Scale). Notice that CanILP chokes when there are $20$ catalogs.}\label{fig:comparison-special-1}
\end{figure}

That being said, this suggests an idea for dividing the whole region of the sky that is of interest into different islands where the sources are clustered together so that instead of solving one big problem, we could break it into smaller problems and make use of parallel computing. Essentially, we first apply a single-linkage clustering algorithm to our dataset, which is done using the DBSCAN algorithm with parameters ``min\_samples" = $2$ and ``eps" = $5 \times \max_{ic \in D}\sigma_{ic}.$ With the chosen parameters, we are essentially performing the friends-of-friends algorithm. It turns out that for our simulation, most of these islands consist of only $1$ source from each catalog. Hence, from now on, we will show the result for this scenario of having $1$ source from each catalog. This situation is not peculiar to our simulation but is, in fact, observed in real data sets from multiple visits of the same part of the sky by the same telescope. Analysis for the multiple sources per catalog will be discussed later.  
As can be seen in Figure \ref{fig:comparison-special-1}, even though we are able to handle more than $3$ catalogs, the maximum number of catalogs we could analyze in a day is $20.$ Though, similar to \citet{shi2019probabilistic}, we do not include any pruning procedures, such as those described in \citet{kunszt2001}; \citet{Gorski_2005}; \citet{gray2007}; \citet{lee2017}. These pruning procedures might speed up the matching, but from our experience the complexity of the problem is still exponential in terms of the number of catalogs. The next paragraph discusses how far we could get using DirILP formulation.

\paragraph{DirILP formulation analysis.}
The main drawback from the previous approach is that the process of creating potential subsets $T$ is exponential in terms of the number of catalogs. Even if we consider the island, the number of nonempty subsets in such an island will still be $2^C - 1,$ so creating the variables for the ILP takes a tremendous amount of time.

\begin{figure}[ht]
    \centering
    \includegraphics[width=0.48\textwidth]{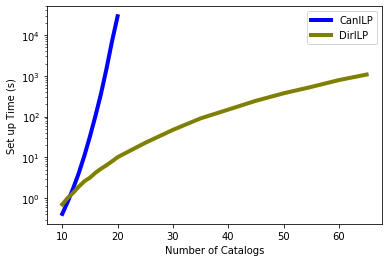}
    \caption{Set up time comparison between the two formulations for the special case (Log Scale)}\label{fig:comparison-special-2}
\end{figure}

\begin{figure}[ht]
    \centering
    \includegraphics[width=0.48\textwidth]{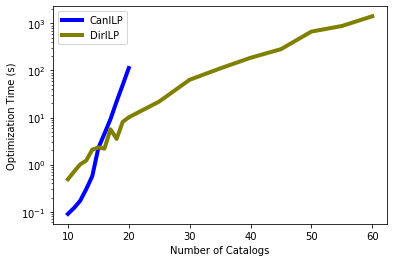}
    \caption{Optimization time comparison between the two formulations for the special case (Log Scale)}\label{fig:comparison-special-3}
\end{figure}

DirILP formulation attempts to fix that problem by reducing the time complexity to create the variables for the ILP to something that is polynomial in the total number of sources. However, since this catalog matching problem is intrinsically difficult, we still have to tackle the exponential complexity of the problem somewhere else: this appears in the time needed to solve the ILP. We believe that with advances in the field of integer linear programming, the Gurobi solver will be able to solve this problem more efficiently. It turns out using DirILP, we are able to tackle up to $60$ catalogs. The comparison for the total running time between CanILP and DirILP is shown in Figure \ref{fig:comparison-special-1}. In addition, we also include the set up time and optimization time for each formulation in Figures \ref{fig:comparison-special-2} and \ref{fig:comparison-special-3}.

Moreover, by including some heuristic constraints, such as imposing a time limit between incumbent solutions, on the Gurobi solver, we are able to push the DirILP further to handle $160$ catalogs. 

Finally, it is important to note that the associations given by CanILP and DirILP are the same and they match the ground truth perfectly. Hence, there is no difference in the accuracy of the matching between the two approaches. They only differ in their running time.  

\subsection{General case: different $\kappa_{ic}$ for every detection}

\begin{figure}[ht]
    \centering
    \includegraphics[width=0.48\textwidth]{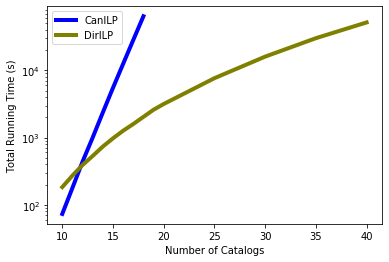}
    \caption{Total running time comparison between the two formulations for the general case (Log Scale). Notice that CanILP chokes when there are $18$ catalogs.}\label{fig:comparison-general-1}
\end{figure}

\begin{figure}[ht]
    \centering
    \includegraphics[width=0.48\textwidth]{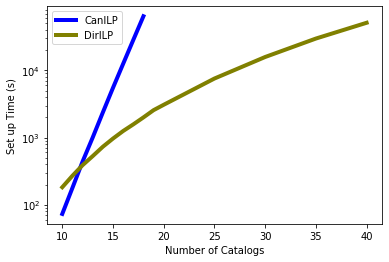}
    \caption{Set up time comparison between the two formulations for the general case (Log Scale)}\label{fig:comparison-general-2}
\end{figure}

\begin{figure}[ht]
    \centering
    \includegraphics[width=0.48\textwidth]{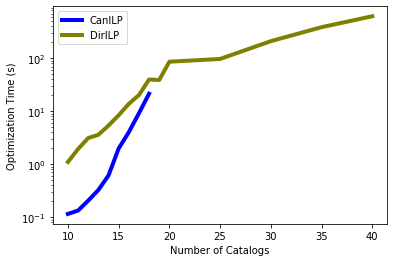}
    \caption{Optimization time comparison between the two formulations for the general case (Log Scale) }\label{fig:comparison-general-3}
\end{figure}

For the general case, both approaches still give all correct associations that match the ground truth. However, as in the special case, DirILP is still more efficient at solving the matching problem than CanILP, as shown in Figure \ref{fig:comparison-general-1}. We should point out that even though in this general setting, the optimal value found in DirILP is just an approximation of the Bayes factor associated with the ground-truth matching, the values are still quite close to each other. More importantly, the associations obtained from DirILP still match the ground-truth associations. Figures \ref{fig:comparison-general-1} - \ref{fig:comparison-general-3} show the total running time, time to set up the ILP, and time for Gurobi to solve the ILP, for both CanILP and DirILP in this general case.

\subsection{Multiple sources per catalog in each island}
\begin{figure}[ht]
    \centering
    \includegraphics[width=0.48\textwidth]{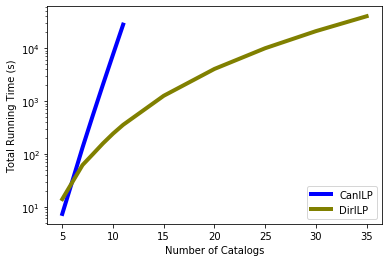}
    \caption{Total Running time comparison between the two formulations when there are $2$ detections from each catalog in an island (Log Scale)}\label{fig:2-detection}
\end{figure}

Recall that in the previous sections, we assume that in each island there is only one detection from each catalog, which is a reasonable assumption in many real-life situations. In this section, we would like to discuss scenarios when the uncertainty $\sigma_{ic}$ is large or the source density is very high. These scenarios will result in islands where there might be multiple detections from each catalog in an island. It turns out that in our simulation, CanILP and DirILP still give the correct association under this scenario. However, both methods run much slower than in the previous scenario and are able to handle only about half as many catalogs with the same settings on the algorithms. We give an example of the running time for the $2$ methods when there are $2$ detections from each catalog. One can see how much worse it can get when the number of detections from each catalog becomes larger. Figure \ref{fig:2-detection} shows the total running time for both CanILP and DirILP when there are $2$ detections from each catalog in an island. 

\subsection{Discussion of running time complexity}\label{sec:summary}
We now give a brief explanation for the shape of the curves in Figures \ref{fig:comparison-special-1}--\ref{fig:comparison-general-3}. For CanILP, since the number of variables is exponential in terms of the number of catalogs, under the log scale as in Figures \ref{fig:comparison-special-2} and \ref{fig:comparison-general-2}, the time to create these variables and set up the ILP as a function of the number of catalogs is represented by a straight line. On the other hand, for DirILP, we have a curve with decreasing gradient instead of a straight line because the number of variables and constraints in this method is polynomial in the number of catalogs. The explanation for the curves in Figures \ref{fig:comparison-special-3} and \ref{fig:comparison-general-3} are similar because the amount of time to solve an ILP generally depends on the number of variables and constraints in the problem. That being said, the curves in these two figures look more jagged because of other complexities involved in the optimization procedure. Finally, as most of the time to solve the catalog matching problem is spent on setting up the ILP, the curves in Figures \ref{fig:comparison-special-1} and \ref{fig:comparison-general-1} are very much similar to their counterparts in Figures \ref{fig:comparison-special-2} and \ref{fig:comparison-general-2}, respectively. 

\section{Implementation and software}
\label{sec:software}
CanILP and DirILP algorithms are implemented in several Jupyter notebooks. They share a common structure: In the first part, we create a simulation with different catalogs and a number of mock objects on each of the catalogs. Next, we perform the DBSCAN algorithm to output different islands, or clusters of detections. Again, as mentioned in \ref{subsec:Homoscedasticity}, with our chosen parameters, this is similar to executing a friends-of-friends algorithm. The reason we pick DBSCAN is because of its well-developed library in Python.

After running the clustering algorithm, we implement CanILP and DirILP to solve the catalog matching problem in each island. The optimization problems in these modules were solved using Gurobi software \citet{gurobi}.

In addition, we employ several (optional) heuristics in the DirILP algorithm to speed up the catalog matching procedure. 
First, any $2$ sources that are more than $8\sigma$ away from each other, we force them to belong to separate objects. 
Second, for sources that are $0.1\sigma$ away from each other, we restrict them to belong to the same object. Finally, we set an MIP Gap (optimality gap) of $0.5\%$ to prevent Gurobi from taking too long to verify optimality. 
Through our experiments, we have found that running the algorithm with these heuristics give the same results but it was 10 times faster. 
The notebooks can be found on Github at  \url{https://github.com/tunguyen52/Nway-matching}.

\section{Summary and Future Work}\label{sec:future}
We have shown how the CanILP approach of \citet{shi2019probabilistic} and the new DirILP solve for a globally optimal matched catalog in crowded fields where a greedy approach is not sufficient. 
The former enumerates the candidate associations and picks the optimal combination of those; the latter introduces variables to directly assign sources to objects.
The new DirILP formulation is superior in the sense that it scales to large number of catalogs better, i.e., produces results in less time. 
The method comes at a price, which is in complexity of the algorithm, especially in case of heteroscedasticity when the catalogs have different astrometric uncertainty. 
In fact DirILP only out-performs the previous method when many catalogs are to be crossmatched.
We recommend the simpler CanILP approach for small number of catalogs, where the combinatorial explosion is not as severe. 
In our experiments, this crossover threshold appears to be at around 12 visits or catalogs, beyond which the DirILP method gets faster.

Both of these methods optimize the same objective and yield the best possible catalog matching result in a likelihood sense. No prior on the partition is imposed currently in our study and the accompanying software. While placing priors on the partition might seem complicated, certain simple priors can be easily expressed, such as those that depend only the number of objects in the matched catalog. This is a possible direction to explore in the future.

Additional future work includes testing the software and its performance on imaging surveys, such as multiple visits of the Hyper Suprime-Cam Subaru Strategic Program, whose data collection resembles future observations of the LSST. 

\begin{acknowledgements}
This material is based upon work supported by the National Science Foundation under Grant No.~1909709, 1814778, 1452820 and 1934979. 
T.B. gratefully acknowledges the aforementioned AST/AAG grants from NSF and funding from NASA via awards STScI-49721 and STScI-52333 under NAS5-26555. 
T.N. and A.B. gratefully acknowledge support from the aforementioned NSF grant and ONR grant N000141812096. 
A.B. is also grateful for support from the aforementioned NSF grant and AFOSR grant FA95502010341. 
The authors thank the anonymous referee for the careful review and the thoughtful comments.
\end{acknowledgements}

\appendix 

\section{DirILP Formulation - Special Case}
\label{appx:DirILP_special}
The ILP Formulation for DirILP when $\kappa_{ic} = \frac{1}{\sigma^2}$ for every source $(i,c)$ is given below.

The objective function we want to minimize is given by
\begin{equation} \min \sum_o \bigg(\;\;p^o + \sum_p a_p w_p^o + t^o \bigg) \end{equation}

The following constraints restrict $x_{ic}^o, y_{ic, i'c'}^o, z_k^o, w_p^o$ to binary variables and $t^o$ to have non-negative values.
\begin{equation} x_{ic}^o, y_{ic, i'c'}^o, z_k^o, w_p^o \in \mathbb{Z}  \text{ and }  0 \leq x_{ic}^o, y_{ic, i'c'}^o, z_k^o, w_p^o \leq 1, \; 0 \leq t^o, \; \; \forall (i,c), k, p, o. \end{equation}

The next equation ensures that all sources $(i,c)$ need to belong to exactly one subset:
\begin{equation}\label{eq:DirILP_special_l1}\sum_o x_{ic}^o = 1, \;\;\forall (i,c) \end{equation} 

The following equation imposes that every subset takes no more than $1$ source from each catalog.
\begin{equation}\sum_i x_{ic}^o \leq 1, \;\;\forall o \in \{1, 2, ..., N\},\;\; \forall c \in \{1, \ldots, C\} \end{equation} \label{eq:DirILP_special_l2}

The following set of constraints on $y_{ic, i'c'}^o$ is an implementation of the definition of $y_{ic, i'c'}^o$ in Section~\ref{sec:DirILP_setup}, which requires $y_{ic, i'c'}^o = 1$ only if $x^o_{ic} = x^o_{i'c'} = 1$:
\begin{gather}\label{eq:DirILP_special_l3} y_{ic, i'c'}^o \geq \; x^o_{ic} + x^o_{i'c'} - 1, \\
y_{ic, i'c'}^o \leq x_{ic}^o, \\
y_{ic, i'c'}^o \leq x_{i'c'}^o,
\end{gather}
for all $(i,c) \neq (i',c') \text{ and } \forall o$.

Since the cardinality of any subset from a partition $P$ is between $0$ and $C,$ the following equation states that only $1$ of $z_k^o$ can take a value of $1$. 
\begin{equation}\sum_{k=0}^C z_k^o = 1, \forall o, \label{eq:DirILP_special_l4} \end{equation}

The next constraint is the definition of $w_p^o$ as described in Section~\ref{sec:DirILP_setup}.
\begin{equation} w_1^o \geq w_2^o \geq \cdots \geq w^o_C \text{ and } \sum_{p=1}^C w_p^o = \sum_{ic} x_{ic}^o, \;\; \forall o, \label{eq:DirILP_special_l5} \end{equation}

With the specific choice of the constant $M$ as defined below, the equation that follows becomes redundant when $z^o_k = 0$ since the RHS will be negative and so $t_o \geq 0$ becomes the enforcing constraint, and when $z^o_k = 1$, the minimization forces $t^o$ to be equal to the first term of the RHS.
\begin{equation}t^o \geq \frac{\sum\kappa\psi_{ic,i'c'}^2 y_{ic, i'c'}^o}{4k} - (1 - z_k^o) M,\;\; \forall o \text{ and } k \in \{1, 2, \cdots, C\}, \label{eq:DirILP_special_l6} \end{equation}
where $M = \bigg\lceil \sum\limits_{ic, i'c' \in D}\frac{\kappa\psi^2_{ic,i'c'}}{4} \bigg\rceil$.

The following set of equations constitutes the definition of $z_k^o.$
\begin{gather} \sum_{ic}x_{ic}^o \leq k z_k^o + C(1 - z_k^o)\label{eq:DirILP_special_l7}\\
\sum_{ic}x_{ic}^o \geq k z_k^o,\label{eq:DirILP_special_l7-II}
\end{gather}
for all $k \in \{0, 1, 2, \cdots, C\}$ and for all $o$.

Finally, the last equation \begin{equation}
p^o \geq \ln(2\kappa)(1 - \sum_p w_p^o) - \ln(2\kappa)z_0^o, \;\; \forall o \label{eq:DirILP_special_l8},
\end{equation}
ensures that for an empty subset $S_o$, $p^o = 0$, hence contributing nothing to the objective. This is because when $z_0^o = 1$ (nothing is assigned to subset $S_o$), $w_p^o = 0, \forall p$. As we are minimizing the objective function with respect to $p^o$, $p^o$ will be set to $0$. On the other hand, when $z_0^o = 0$, the constraint becomes 
$p^o \geq \ln(2\kappa)(1 - \sum_p w_p^o)$ and again, since we are minimizing, $p^o$ will equal this value.

\section{DirILP Formulation - General Case}
\label{appx:DirILP_general}

Below, we give the ILP Formulation for DirILP when $\kappa_{ic}$ is different for distinct sources $(i,c)$. Some of these constraints are similar to the special case so we will only give explanations for the new constraints, which are shown after the ILP formulation. We follow the notation introduced in Section~\ref{sec:DirILP_general}. In particular, recall the constants $c_0, c_1, \ldots, c_Q$ designed to model, up to the nearest 100, the sum of subsets of uncertainties $\kappa_{ic}$, the associated decision variables $u^o_k$ and the decision variables $\chi^o_P$ to model the logarithms of such sums.

The objective in the general case is
\begin{equation}\label{eq:obj-general}
\min \sum_o \bigg(\;\;p^o - \sum_{ic}x^o_{ic}\ln\kappa_{ic} + \chi_1^o b_{\min} + \epsilon\sum_{p=2}^R \chi_p^o + t^o \bigg)
\end{equation}

As in the special case, the following constraints on the variables restrict $x_{ic}^o, y_{ic, i'c'}^o, z_k^o$ to binary variables and $t^o$ to have non-negative values. Additionally, the new variables $\chi_p^o, u^o_k$ are also restricted to binary values.

$$x_{ic}^o, y_{ic, i'c'}^o, z_k^o, \chi_p^o, u^o_k \in \mathbb{Z}  \text{ and }  0 \leq x_{ic}^o, y_{ic, i'c'}^o, z_k^o, \chi_p^o, u^o_k \leq 1, \;\; 0 \leq t^o.$$
\medskip

Next, constraints~\eqref{eq:DirILP_special_l1}--\eqref{eq:DirILP_special_l4} and~\eqref{eq:DirILP_special_l7}--\eqref{eq:DirILP_special_l7-II} are included verbatim. 
\medskip

The following impose the conditions required on $\chi^o_p$ as described in Section~\ref{sec:DirILP_general}. 

\begin{equation}\label{eq:DirILP_general_l2} \chi_1^o \geq \chi_2^o \geq \cdots \geq \chi^o_R \text{ and } \chi_1^o \exp(b_1) + \sum_{p=2}^R \chi_p^o (\exp(b_p) - \exp(b_{p-1})) \geq  \sum_{ic} \kappa_{ic} x_{ic}^o, \;\; \forall o.
\end{equation}
\medskip

The next set of constraints ensure that the value of $\sum_{ic \in S_o} \kappa_{ic}$ will be approximately equal to $c_k$ (up to the nearest 100) for some $k \in \{0, 1, 2, \cdots Q\}.$

\begin{equation}
    \sum_{k=0}^Q u_k^o = 1, \;\;\forall o,\label{eq:DirILP_general_l1}
\end{equation}
\begin{equation}\label{eq:DirILP_general_l3}
    \begin{array}{l} \sum_{ic}[\kappa_{ic}]_{_{100}}x_{ic}^o \leq c_k u_k^o + M'(1 - u_k^o)\\
\sum_{ic}[\kappa_{ic}]_{_{100}}x_{ic}^o \geq c_k u_k^o \end{array}, \;\;\forall k \in \{0, 1, 2, \cdots, Q\} \text{ and } \forall o.
\end{equation}
where $M' = C\max_{ic \in D}\kappa_{ic}.$

Finally, the last set of constraints tie everything back into the objective function~\eqref{eq:obj-general}.

\begin{gather}
    t^o \geq \frac{\sum_{ic}\sum_{i'c'}\kappa_{ic}\kappa_{i'c'}\psi_{ic,i'c'}^2 y_{ic, i'c'}^o}{4c_k} - (1 - u_k^o) M,\;\; \forall o \text{ and } k \in \{1, 2, \cdots, Q\}, \\
    p^o \geq (1 - \sum_{ic \in S_o} x_{ic}^o)\ln2 - z_0^o\ln2, \;\; \forall o.
\end{gather}
where $M = \Bigg\lceil \frac{\max_{ic \in D}\kappa_{ic}^2\sum\limits_{ic,i'c' \in D} \psi_{ic,i'c'}^2}{4\min_{ic \in D} \kappa_{ic}} \Bigg\rceil$.

\bibliography{ref.bib}

\end{document}